\documentclass[
 reprint,
 superscriptaddress,
 amsmath,amssymb,
 aps,
 prb,
]{revtex4-2}

\usepackage{graphicx}
\usepackage{dcolumn}
\usepackage{bm}
\usepackage{xcolor}
\usepackage{hyperref}

\begin{document}

\title{Exceptional thermoelectric properties in Na$_2$TlSb enabled by quasi-1D band structure}

\author{Øven A. Grimenes}
\email{oven.andreas.grimenes@nmbu.no}
\affiliation{
 Department of Mechanical Engineering and Technology Management, \\ Norwegian University of Life Sciences, NO-1432 Ås, Norway}

\author{Ole M. Løvvik}
\affiliation{SINTEF Sustainable Energy Technology, Forskningsveien 1, NO-0314 Oslo, Norway}

\author{Kristian Berland}
\email{kristian.berland@nmbu.no}
\affiliation{
 Department of Mechanical Engineering and Technology Management, \\ Norwegian University of Life Sciences, NO-1432 Ås, Norway}

\date{\today}

\begin{abstract}
\noindent
Materials with reduced dimensionality offer beneficial density-of-states (DOS) profiles for thermoelectric energy conversion, but can be impractical in realistic devices. Encouragingly, bulk high-symmetry materials can also exhibit similar quasi-low-dimensional band structures. A striking example is the full-Heusler compound Na$_2$TlSb, whose valence-band energy isosurfaces can form intersecting two-dimensional pockets, i.e., a box-like structure. The individual energy isosurface sheets resemble those  of 1D quantum wires. The combination of high electron velocities (perpendicular to the pockets) and a rapidly increasing DOS with energy in the transport regime (due to the low dimensionality) makes Na$_2$TlSb a representative case where the band structure gives rise to attractive electronic transport properties. However, these beneficial features could be counteracted by high electronic scattering rates due to the large scattering space. In this first principles study of Na$_2$TlSb we find that the electronic scattering rates remain modest. This result is linked to the reduced matrix elements of large-momentum ($\mathbf{q}$) scattering across the delocalized energy isosurfaces. The enhanced free-carrier screening due to the large DOS also contributes to reducing scattering. In combination, the low-dimensional features and modest scattering result in excellent electronic transport properties. Combined with an ultra-low lattice thermal conductivity of $\kappa_\ell < 1$\;W/mK reported in the literature, we predict a thermoelectric figure of merit ranging from 2.4 at 300\;K to a 4.4 at 600\;K. The $n$-type properties are also favorable, with $zT$ values from 1.5 at 300\;K to 3.0 at 600\;K.
\end{abstract}

\maketitle

\section{Introduction}

% General TE
Thermoelectric (TE) materials are widely used in niche applications for cooling and electricity generation due to their high reliability, noiseless operation, and modular scalability \cite{Bell_2008}. However, their efficiency remains significantly lower than that of conventional vapor-liquid systems \cite{Snyder_2008,Schwab_2022,Bell_2008}. The efficiency of a TE material at a temperature $T$ is linked to the TE figure of merit $zT = S^2\sigma T/(\kappa_\mathrm{e} + \kappa_\ell)$, where $S$ is the Seebeck coefficient and $\sigma$ is the electrical conductivity, while $\kappa_\mathrm{e}$ and $\kappa_\ell$ are the electronic and lattice parts of the thermal conductivity. These properties are highly interdependent, and increasing $zT$ is non-trivial. In particular, strategies that enhance the density of states near the band edge can increase the number of available scattering processes.

Reducing $\kappa_\ell$ by e.g., nanostructuring \cite{Minnich_2009}, nanoinclusions \cite{Theja_2022}, and isovalent alloying \cite{Heinrich_2014,Eliassen_2017,Tranas_2022} has seen success across many systems, but can also have negative effects on the electronic transport properties. On the electronic side, a higher $zT$ can be achieved by increasing the carrier mobility \cite{Wang_2023a}, by introducing nanostructures that can lead to energy filtering \cite{Faleev_2008,Gayner_2020,Lin_2020}, or enhancing the electron density of states (DOS) close to the band edge \cite{Heremans_2008}. An increased DOS can be achieved by aligning bands or valleys through strain or alloying \cite{Zhang_2016,Li_2018,Pei_2011,Zhu_2022} or by introducing resonant states \cite{Bilc_2004,Heremans_2008,Heremans_2012}.

\begin{figure}[t!]
    \centering
    \includegraphics[width=\linewidth]{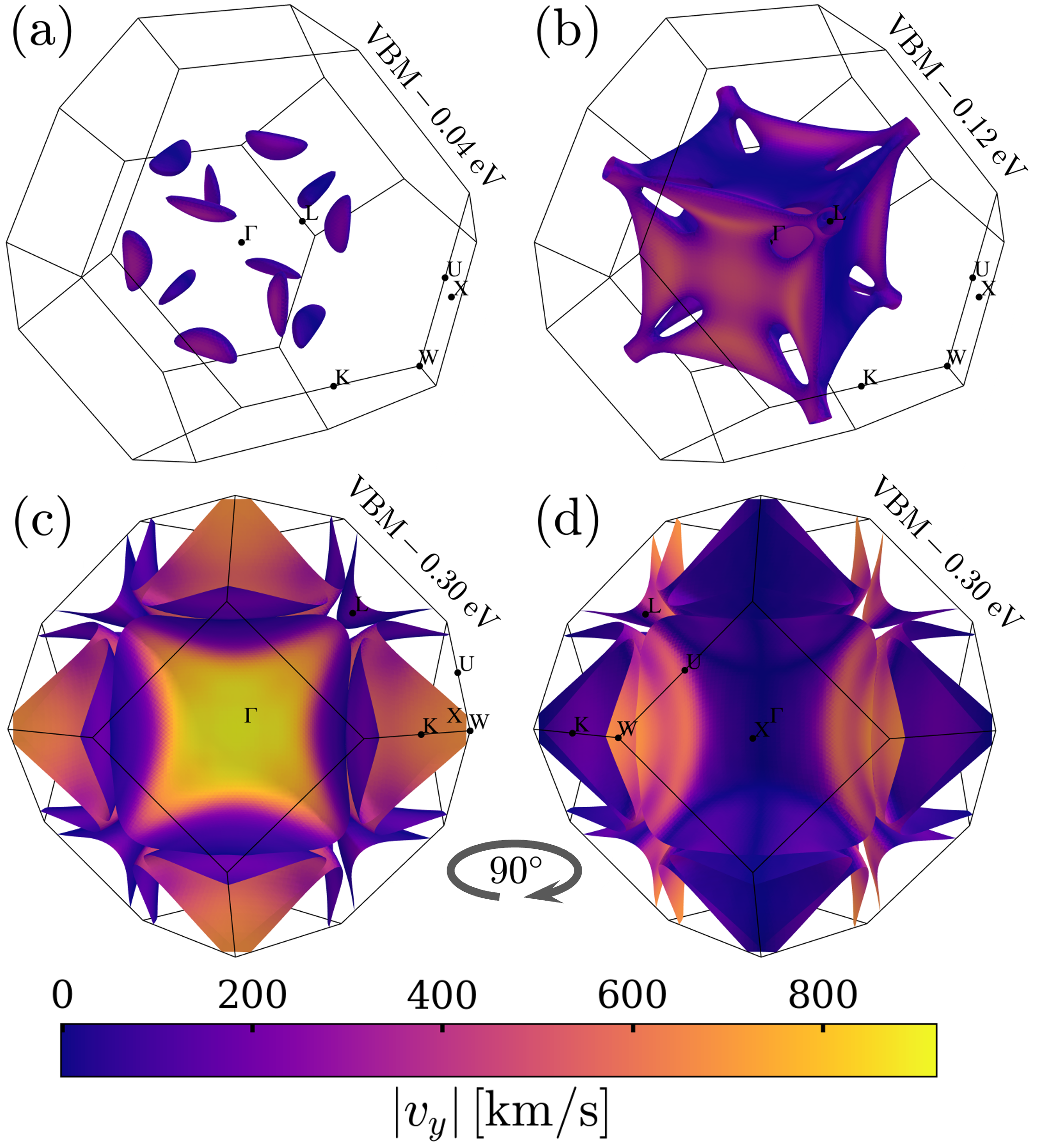}
    \caption{Energy isosurfaces of the Na$_2$TlSb valence band obtained with density functional theory. Just below the band edge (a), the valence band forms 12 Fermi pockets between $\Gamma$ and K, but these merge into continuous box-like isosurfaces shown at 0.12\;eV below the VBM in (b). The box surfaces consist of pairs of sheets giving a hollow energy isosurface. Moving to  0.30\;eV below the VBM, the isosurface expands to six overlapping sheets, shown from two different perspectives in (c) and (d). The color gradient indicates the $y$-component of the electron group velocity, highlighting how the sheets contribute individually to transport in the direction perpendicular to the sheets.
    \label{fig: fermi}}
\end{figure}
% Low-dim
Low-dimensional material structures, e.g., thin layers (2D materials) or nanowires (1D), can give rise to attractive density of state profiles and thus excellent electron transport properties \cite{Hicks_1993,Hicks_1993a}. Such structures can exhibit promising properties, \cite{Boukai_2008,Hochbaum_2008,Venkatasubramanian_2001}, but demands the dimension of the nanostructures to be sufficiently small \cite{Cornett_2011,Cornett_2011a}. Encouragingly, conceptual studies indicate that band structures exhibiting 2D-like (tubes) and 1D-like (sheets) energy isosurfacess can arise even in bulk, isotropic materials \cite{Parker_2013,Bilc_2015,Dylla_2019,Park_2021,Brod_2021}. This was also confirmed in our recent study of the TE properties of  CsK$_2$Sb,  which hosted extended tube-like energy isosurfaces near the valence band edge \cite{Grimenes_2025}. PbTe \cite{Brod_2020} also hosts extended sheet-like energy isosurfaces within the valence band. As does the related full-Heusler compounds Li$_2$TlBi and Li$_2$InBi \cite{He_2019}, but in these cases they are located too far below the valence band maximum (VBM) to contribute considerably to the electron transport. 

% Na2TlSb
A system with extended energy isosurfaces close to the VBM is the full-Heusler Na$_2$TlSb (see Fig.~\ref{fig: fermi}). This material has previously been studied by Yue \textit{et al.} \cite{Yue_2023}, where the ultralow $\kappa_\ell$ was identified as the primary origin of its high predicted $zT$. However, the electronic transport properties and, in particular, the nature of the electron scattering associated with its quasi-low-dimensional valence band were not analyzed in detail. The system itself was independently identified as a material with very low $\kappa_\ell$ by us \cite{Tranas_2023a}. In the current study, we investigate the TE potential of Na$_2$TlSb with a focus on the interplay between the electronic band structure and the electronic scattering. Analyzing the momentum-dependent scattering mechanisms, we show that the large density of states associated with sheet-like isosurfaces does not necessarily lead to enhanced effective momentum relaxation, due to a combination of the momentum dependence of the scattering elements and free-carrier screening.

\section{Theory}
\subsection{Electron transport}
Electron transport properties were modeled with the Boltzmann transport equation in the relaxation time approximation \cite{Ziman_1960,Madsen_2018,Ganose_2021}. The electron transport properties can be expressed with the generalized transport coefficients
\begin{equation}
    \label{eq: general_transport_coeff}
    \mathcal{L}^{(\alpha)} = q^2 \int_{-\infty}^{\infty}  \mathrm{d}\varepsilon\,\Sigma(\varepsilon) W^{(\alpha)} (\varepsilon)\,,
\end{equation}
where $q$ is the elementary charge and $W^{(\alpha)}$ are Fermi selection functions, 
\begin{equation}
    \label{eq: self_func}
    W^{(\alpha)}(\varepsilon) = (\varepsilon - \varepsilon_\mathrm{F})^\alpha \left( \frac{\partial f}{\partial \varepsilon} \right) \,.
\end{equation}
Here, $\varepsilon_\mathrm{F}$ is the Fermi level and $f$ is the Fermi-Dirac distribution function. The transport distribution function is given by 
\begin{equation}
    \label{eq: transport_dist}
    \Sigma(\varepsilon) =  \sum_{n} \int_{\mathrm{BZ}} \frac{\mathrm{d}\mathbf{k}}{8\pi^3}\, \tau_{n\mathbf{k}}\, v_{n\mathbf{k}}\otimes v_{n\mathbf{k}} \, \delta(\varepsilon - \varepsilon_{n\mathbf{k}}) \,,
\end{equation}
where $v_{n\mathbf{k}}$ and $\tau_{n\mathbf{k}}$  are the group velocity and relaxation time of a state with index $n\mathbf{k}$. 

In terms of the transport coefficients, the electronic TE properties $\sigma$, $S$, and $\kappa_\mathrm{e}$ can, at the temperature $T$, be expressed as
\begin{align}
    &\sigma = \mathcal{L}^{(0)} \,, \label{eq: conductivity}   \\
    &S = \frac{\mathcal{L}^{(1)}}{q T \mathcal{L}^{(0)}} \,, \label{eq: Seebeck} \\
    &\kappa_\mathrm{e} = \frac{1}{q^2 T}\left[\mathcal{L}^{(2)} - \frac{(\mathcal{L}^{(1)})^2}{\mathcal{L}^{(0)}}\right] \,. \label{eq: kappa_e} 
\end{align}

\subsection{Electron scattering}
The transition rate between two states is given by  
\begin{equation}
    \Gamma_{n\mathbf{k} \rightarrow m\mathbf{k}+\mathbf{q}} = \frac{2\pi}{\hbar} |g_{nm}(\mathbf{k},\mathbf{q})|^2 \delta(\varepsilon_{n\mathbf{k}} - \varepsilon_{m\mathbf{k}+\mathbf{q}})\,,
\end{equation}
where $g_{nm}(\mathbf{k},\mathbf{q})$ are effective coupling matrix elements. In the momentum relaxation time approximation (MRTA) \cite{Li_2015,Ponce_2020,Ganose_2021},
used to compute elastic electron relaxation rates,
the effective relaxation rate of a state is given by
\begin{equation}
    \label{eq: MRTA}
    \Gamma_{n\mathbf{k}}^\mathrm{MRTA} = \sum_m \int  \frac{\mathrm{d}\mathbf{q}}{\Omega_\mathrm{BZ}}\, \Lambda_{nm}(\mathbf{k},\mathbf{q}) \Gamma_{n\mathbf{k} \rightarrow m\mathbf{k}+\mathbf{q}}\,,
\end{equation}
where $\Omega_{\mathrm{BZ}}$ is the Brillouin-zone volume and the geometric MRTA factor
\begin{equation}
    \label{eq: MRTA_factor}
    \Lambda_{nm}(\mathbf{k},\mathbf{q}) = \left[ 1- \frac{\mathbf{v}_{n\mathbf{k}} \cdot \mathbf{v}_{m\mathbf{k}+\mathbf{q}}}{|\mathbf{v}_{n\mathbf{k}}|^2} \right]
\end{equation}
weights the relative importance of forward- and back-scattering on transport properties. For inelastic scattering, the MRTA does not hold \cite{Ponce_2020,Ganose_2021}, and for such scattering, rates were computed using the self-energy relaxation time approximation (SERTA) \cite{Ponce_2020,Ganose_2021}, i.e., the MRTA geometric factor is set to unity. For polar-optical phonons, transition rates are given by
\begin{align}
    & \Gamma_{n\mathbf{k} \rightarrow m \mathbf{k}+\mathbf{q}} = \frac{2\pi}{\hbar} |g_{nm}(\mathbf{k},\mathbf{q})|^2 \\ \nonumber
    & \times [(n^{\mathrm{BE}}_\mathbf{q} + 1 - f_{m \mathbf{k}+\mathbf{q}}) \delta(\varepsilon_{n\mathbf{k}} - \varepsilon_{m \mathbf{k} + \mathbf{q}} - \hbar \omega_\mathbf{q}) \\ \nonumber
    & + (n^{\mathrm{BE}}_\mathbf{q} + f_{m \mathbf{k} + \mathbf{q}}) \delta(\varepsilon_{n\mathbf{k}} - \varepsilon_{m \mathbf{k} + \mathbf{q}} + \hbar \omega_\mathbf{q})],
\end{align}
where $n^{\mathrm{BE}}_\mathbf{q}$ and $\omega_\mathbf{q}$ are the Bose-Einstein occupancy and frequency of the optical phonon with wave vector $\mathbf{q}$, respectively. The corresponding relaxation rates are given by 
\begin{equation}
    \Gamma_{n\mathbf{k}}^\mathrm{SERTA} = \sum_m \int \frac{\mathrm{d}\mathbf{q}}{\Omega_\mathrm{BZ}} \Gamma_{n\mathbf{k} \rightarrow m \mathbf{k}+\mathbf{q}}.
\end{equation}

Within \textsc{AMSET} \cite{Ganose_2021}, which was used in this study, the coupling terms involved in the scattering mechanisms are approximated as a product of a perturbing potential and the wavefunction overlap between the initial and final state,
\begin{equation}
    M_{mn}(\mathbf{k},\mathbf{q}) = \langle m \mathbf{k}+\mathbf{q}|e^{i\mathbf{q}\cdot \mathbf{r}}| n\mathbf{k}\rangle\, .\label{eq: M}
\end{equation}
Both $M$ and the perturbing potential depend on the momentum transfer $\mathbf{q}$. The perturbative potential is given in terms of effective coupling elements for the acoustic deformation potential (ADP), ionized impurity (IMP), and polar optical phonon (POP) scattering, which can be expressed as
\begin{align}
    &g_{nm}  ^{\mathrm{ADP}} (\mathbf{k},\mathbf{q}) = A^\mathrm{ADP}  M_{mn}(\mathbf{k},\mathbf{q}) \;,  \label{eq: ADP1} \\
    &g_{nm}  ^{\mathrm{POP}} (\mathbf{k},\mathbf{q}) = A^\mathrm{POP} \frac{M_{mn}(\mathbf{k},\mathbf{q})}{\sqrt{\mathbf{q}^2 + \beta_\mathrm{\infty}^2}}\;,\label{eq: POP1}  \\
    &g_{nm}  ^{\mathrm{IMP}} (\mathbf{k},\mathbf{q}) = A^\mathrm{IMP} \frac{   M_{mn}(\mathbf{k},\mathbf{q})}{\mathbf{q}^2 + \beta_\mathrm{s}^2}\;,\label{eq: IMP1}
\end{align}
where the prefactors $A$ are independent of the amplitude of $q$. The high-frequency inverse free-carrier screening length is given by
\begin{equation}
    \label{eq: screening}
    \beta^2_{\infty} = \frac{e^2}{\mathbf{\epsilon}_{\infty} k_\mathrm{B}T}\int \frac{\mathrm{d}\varepsilon}{V}\, \mathrm{DOS}(\varepsilon)f(\varepsilon)(1-f(\varepsilon))\,,
\end{equation}
while the static inverse free-carrier screening length $\beta_{0}$ is calculated the same way, but uses the static dielectric constant $\mathbf{\epsilon}_\mathrm{s}$ in place of the high-frequency $\mathbf{\epsilon}_\infty$. 

The prefactor in the perturbing potential of ADP scattering is given by
\begin{equation}
    \label{eq: ADP}
    A^\mathrm{ADP} = \sqrt{k_\mathrm{B} T} \sum_{i = l,t_1,t_2} \frac{\mathbf{\tilde{D}}_{n\mathbf{k}} : \mathbf{\hat{S}}_i}{c_i\sqrt{\rho}}\,,
\end{equation}
where $\mathbf{\tilde{D}}_{n\mathbf{k}}$ is the velocity-corrected deformation potential $\mathbf{D}_{n\mathbf{k}} + v_{n\mathbf{k}} \otimes v_{n\mathbf{k}}$. Here '$:$'  denotes the double inner product, $\mathbf{\hat{S}}$ is the unit strain, $c$ is the sound velocity, and $\rho$ is the mass density. The sum runs over the longitudinal $l$ and the two transverse $t_1$, $t_2$ directions for $\mathbf{\hat{S}}$ and $c$.

For IMP scattering, the prefactor is given by
\begin{equation}
    \label{eq: IMP}
    A^{\mathrm{IMP}} = \frac{n_{ii}^{1/2} Z e}{\hat{\mathbf{n}} \cdot \mathbf{\epsilon}_s \cdot \mathbf{\hat{n}}}\,,
\end{equation}
where $n_{ii} = (n_\mathrm{h} - n_\mathrm{e})/Z$ is the concentration of ionized impurities, $Z$ is the impurity charge state, and $\mathbf{\hat{n}}$ is a unit vector in the direction of scattering, 

For POP scattering,
\begin{equation}
    \label{eq: POP}
    A^\mathrm{POP} = \left[ \frac{\hbar \omega_{\mathrm{po}}}{2} \right]^{1/2} \left( \frac{1}{\mathbf{\hat{n}} \cdot \mathbf{\epsilon}_\infty \cdot \mathbf{\hat{n}}} - \frac{1}{\mathbf{\hat{n}} \cdot \mathbf{\epsilon}_\mathrm{s} \cdot \mathbf{\hat{n}}} \right)^{1/2}\,.
\end{equation}
Here,  $\omega_\mathrm{po}$ is an effective weighted optical phonon frequency given by $\omega_\mathrm{po} = \sum_\nu \omega_{\Gamma \nu} w_\nu / \sum_\nu w_\nu\,,$ where $\omega_{\Gamma \nu}$ is the phonon frequency of branch $\nu$ at $\Gamma$ and $w_\nu = \sum_\gamma \left[ m_\gamma \omega_{\mathbf{q}\nu} \right]^{-1/2} \times [\mathbf{q} \cdot \mathbf{Z}_\gamma ^* \cdot \mathbf{e}_{\gamma \nu}]$, where $m$ is the atomic mass of atom $\gamma$, $\mathbf{Z}_\gamma ^*$ is the Born effective charge, and $\mathbf{e}_{\gamma \nu}$ is a phonon eigenvector.

The total (TOT) scattering rate was obtained according to Matthiessen's rule \cite{Matthiessen_1860}:
\begin{equation}
    \Gamma_{n\mathbf{k}}^\mathrm{TOT} = \Gamma_{n\mathbf{k}}^\mathrm{ADP} + \Gamma_{n\mathbf{k}}^\mathrm{IMP} + \Gamma_{n\mathbf{k}}^\mathrm{POP}\;.
\end{equation}

\subsection{Computational details}
\label{app: computational_detials}
The main electron transport results in this paper were computed using \textsc{AMSET} \cite{Ganose_2021} version 0.5.0. Input electronic band structure and physical properties were calculated using density functional theory (DFT) as implemented in the Vienna Ab initio Simulation Package (\textsc{VASP}) \cite{Kresse_1993,Kresse_1999}. The vdW-DF-cx \cite{Berland_2014,Berland_2014a,Berland_2015} exchange correlation functional was used for relaxing the crystal structure, computing elastic constants with finite differences \cite{LePage_2002}, and the density functional perturbation theory (DFPT) calculations for obtaining the ionic contributions to the dielectric constant and phonon frequencies  \cite{Gajdos_2006}. As vdW-DF-cx only accounts for exchange at the generalized-gradient approximation level, band gaps can be underestimated, the hybrid functional HSE06 \cite{Krukau_2006} was used to calculate the high-frequency dielectric constant, deformation potentials, and the electronic band structure used for transport calculations. A plane-wave cutoff of 520\;eV (identical to the standard of the Materials Project (MP) \cite{Jain_2013} value) was used and the number of electrons in the basis set was 7 (Na), 13 (Tl), and 5 (Sb). DFT calculations included spin-orbit coupling (SOC). The initial crystal structure was retrieved from the MP database (mp-866132) and relaxed to within $5.0\times10^{-5}$\;eV/Å. A second relaxation was subsequently performed to avoid errors related to Pulay stress. Relaxations, DFPT, finite differences for elastic constants, and deformation potentials were calculated using a 12$\times$12$\times$12 $\mathbf{k}$-point grid. The final band structure for transport used a 24$\times$24$\times$24 $\mathbf{k}$-point grid. In the transport calculations, the band structure was interpolated to a 63$\times$63$\times$63 $\mathbf{k}$-point grid. The high-frequency dielectric constant $\epsilon_\infty$ was calculated at the HSE06 level, computing the momentum matrix elements and the electronic levels using a 24$\times$24$\times$24 $\mathbf{k}$-point grid using 80 bands at the independent particle level. We used an in-house code for computing $\epsilon_\infty$ in the independent-particle approximation from the longitudinal matrix elements extracted from \textsc{VASP}. This choice was made since $\epsilon_\infty$ was found to be highly dependent on the band gap and number of $\mathbf{k}$-points. $\epsilon_\infty$ was, on the other hand, quite insensitive to local field corrections, and hence neglected.

For computing $zT$, $\kappa_\ell$ was taken from Ref.\ \cite{Yue_2023} and interpolated using splines, listed in  App. \ref{app: ltc}.

\subsection{\textsc{AMSET} benchmarking and phonon renormalization \label{sec: benchmark}}

To benchmark the electron lifetimes provided by \textsc{AMSET}, we performed supplementary ab initio electron-phonon coupling calculations using recently released \textsc{VASP} routines \cite{Chaput_2019, Engel_2022, Chaput_2025} with the \textsc{Phelel} interface. In the comparison, vdW-DF-cx was used both for \textsc{Phelel} and \textsc{AMSET}. Moreover, to reduce computational costs, the plane-wave energy cutoff was reduced to 400\;eV and less expensive pseudotentials with 1 (Na), 3 (Tl), and 5 (Sb) valence electrons were chosen. The Fan-Migdal self-energy was computed on a $50\times50\times50$ $\mathbf{k}$-point grid and scattering rates were calculated within the self-energy relaxation-time approximation.

The force constants based on the default frozen phonon approximation set up with \textsc{Phelel}, resulted in a phonon dispersion with imaginary frequencies. This is in line with the dynamically unstable structure at 0\;K, also reported in Ref.~\cite{Yue_2023}. To account for thermal renormalization of phonon modes and their effect on scattering rates, we computed force constants with the stochastic temperature-dependent effective potential ({\textsc{sTDEP}}) method \cite{Knoop2024}, using canonical ensembles at $T=600$\;K. Five stochastic configurations with 192 atoms each were used to compute forces with {\textsc{VASP}}, employing a plane-wave cut-off energy of 520\;eV and a $\mathbf{k}$-point density of at least 4 points per reciprocal Å. Spin-orbit coupling was ignored. The extracted force constants were used to generate new generations of configurations until self-consistency was reached with differences in the phonon free energy between consecutive iterations reached less than 1 meV/atom. The cutoff radius of the interatomic interactions in the force extraction procedure was half the minimum super-cell lattice constant (8.25\;{\AA}), giving an overdetermination ratio of 452.

\section{Results}
\subsection{Materials properties}
\label{app: Mat_prop}
Table \ref{tab: Mat_prop} lists the lattice constant,  elastic constants $C_{ij}$, static $\mathbf{\epsilon}_\mathrm{s}$, and high-frequency dielectric constant $\mathbf{\epsilon}_\infty$ obtained for Na$_2$TlSb. The lattice constant of 7.53\;Å is slightly higher than the value of 7.49 Å reported by Yue \textit{et al.}, using PBEsol \cite{Perdew_2008}. The low $C_{ij}$ values result in a low velocity of sound, $c$, in Eq.~\ref{eq: ADP}. The low $c$ can also partly explain the low $\kappa_\ell$ \cite{Isotta_2023}. Low elastic constants can also give rise to high electron scattering rates, as they can increase the ionic polarizability (i.e., giving a large $A^{\mathrm{POP}}$), and hence increase the POP scattering.

\begin{table}[h!]
  \caption{The lattice constant, elastic constants, dielectric tensors, and average optical phonon frequency of Na$_2$TlSb.}
\label{tab: Mat_prop}
\begin{ruledtabular}
\begin{tabular}{lllllll}
$a$ [Å]                     & 7.53      \\
$C_{11}$ [GPa]              & 46.3      \\
$C_{12}$ [GPa]              & 16.8      \\
$C_{44}$ [GPa]              & 24.9      \\
$\epsilon_\mathrm{s}$       & 113.5     \\
$\epsilon_\infty$           & 28.3      \\
$\omega_\mathrm{po}$ [THz]  & 1.55      \\
\end{tabular}
\end{ruledtabular}
\end{table}

\subsection{Electronic structure}
\begin{figure}[t!]
    \centering
    \includegraphics[width=\linewidth]{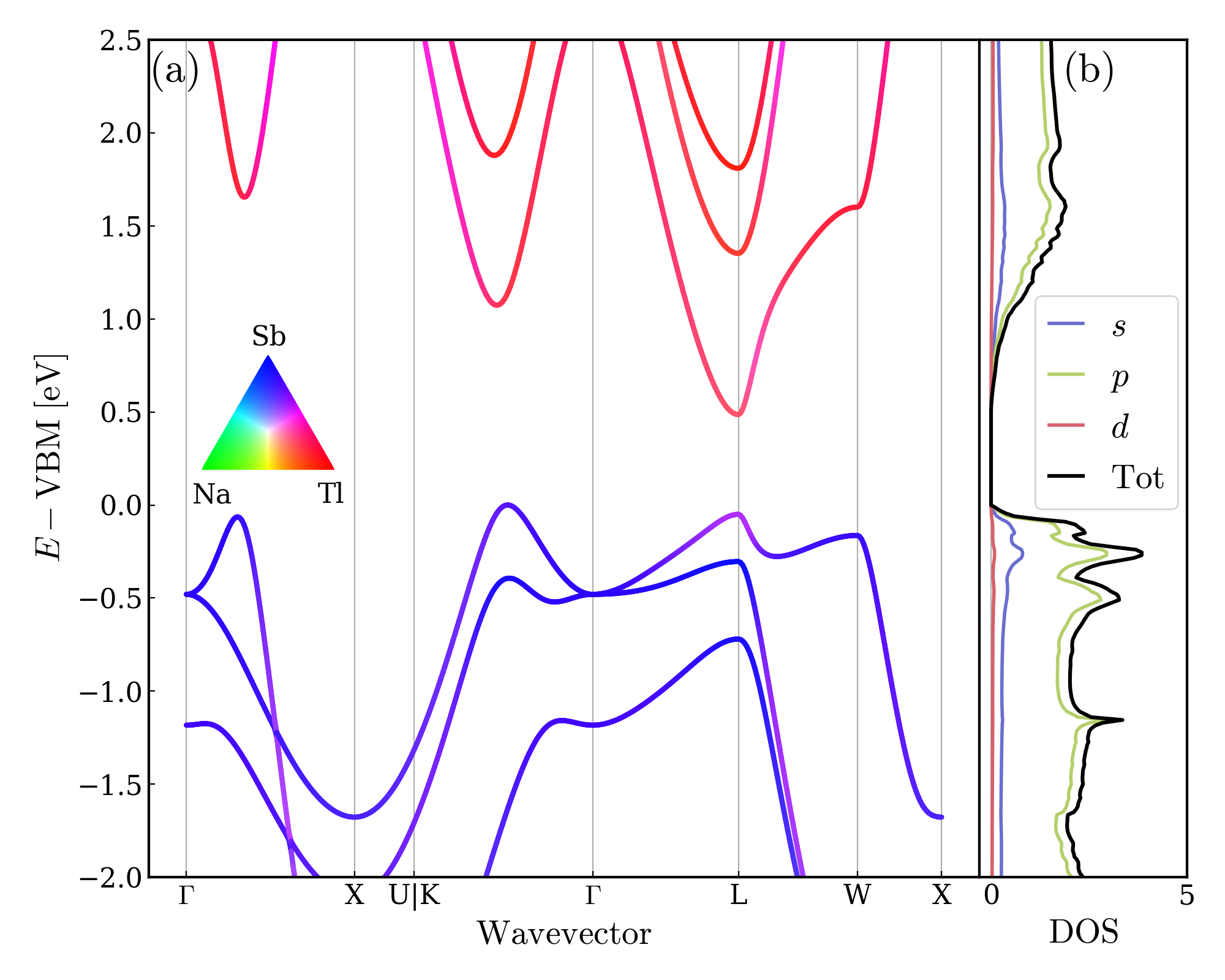}
    \caption{(a) Atom-projected electronic band structure and (b) the orbital and total electronic DOS of Na$_2$TlSb calculated with the HSE06 functional and spin-orbit coupling.
    \label{fig: band_structure}}
\end{figure}

% Band structure and Fermi surface
Figure \ref{fig: band_structure} shows the electronic band structure and density of states (DOS) corresponding to the energy isosurfaces of Fig.~\ref{fig: fermi}, which resembles that of an ideal 1D material \cite{Brod_2021}. The quasi-low-dimensional band structure can explain the rapidly rising valence-band DOS, as compared to the conduction-band DOS. Both the conduction and valence bands are strongly dominated by p-orbital character, mostly associated with Sb for the valence band and Tl for the conduction band. The band character is akin to PbTe and analogous compounds such as Li$_2$TlBi \cite{Brod_2020,He_2019}, where Tl (Sb) takes the role of Pb (Te), with Na as a spectator cation. For a more in-depth analysis of the origin of the quasi-low-dimensional features in PbTe, see Ref. \cite{Brod_2021}. With HSE06, we found a band gap of 0.49\;eV.

\subsection{Electronic transport}
\begin{figure}[h!]
    \centering
    \includegraphics[width=\linewidth]{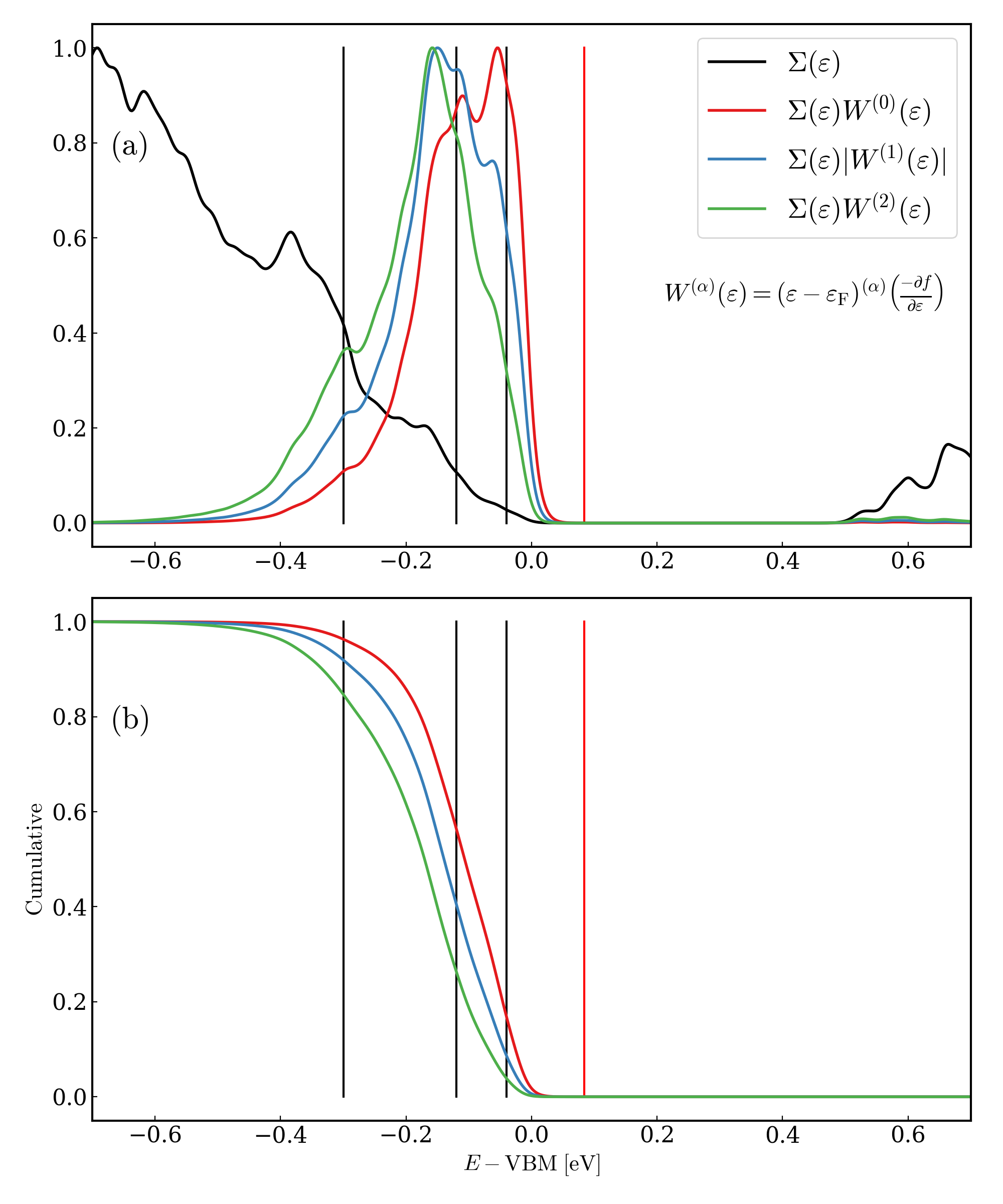}
    \caption{Effective (a) and cumulative (b) normalized transport contributions at different energies, with Fermi selection functions $W^{(\alpha)}$ in Eq. \ref{eq: self_func} based on the Fermi level optimizing $zT$ (red vertical line) at 600\;K. The vertical black lines correspond, respectively, to the energy isosurfaces of Fig. \ref{fig: fermi}.
    \label{fig: vvdos}}
\end{figure}

Figure \ref{fig: vvdos} shows the normalized transport distribution $\Sigma(\varepsilon)$  for respective contributions to transport coefficients at 600\;K, with the Fermi level set to the value optimizing $zT$ ($\varepsilon_\mathrm{F} = 0.084$\;eV). For the product of $\Sigma$ and $W^{(0)}$, which can be used to determine the electrical conductivity, the highest contribution is located at an energy of  $-0.055$\;eV relative to the VBM. This energy is close to the multi-pocket isosurface surface at $-0.04$\;eV shown in Fig. \ref{fig: fermi} (a). However, the cumulative curve in Fig. \ref{fig: vvdos} (b) shows that most of the contribution occurs further into the band, at energies where the isosurface exhibit clear low-dimensional features. For $\Sigma W^{(1)}$ and $\Sigma W^{(2)}$, which have a larger spread in energy, the largest contribution is also further into the valence band. This also shifts the cumulative curves, resulting in the major contributions to $\Sigma W^{(1)}$ and $\Sigma W^{(2)}$ to be in the energy range between the surfaces shown in Fig. \ref{fig: fermi} (b) and (c).

\begin{figure}[h!]
    \centering
    \includegraphics[width=\linewidth]{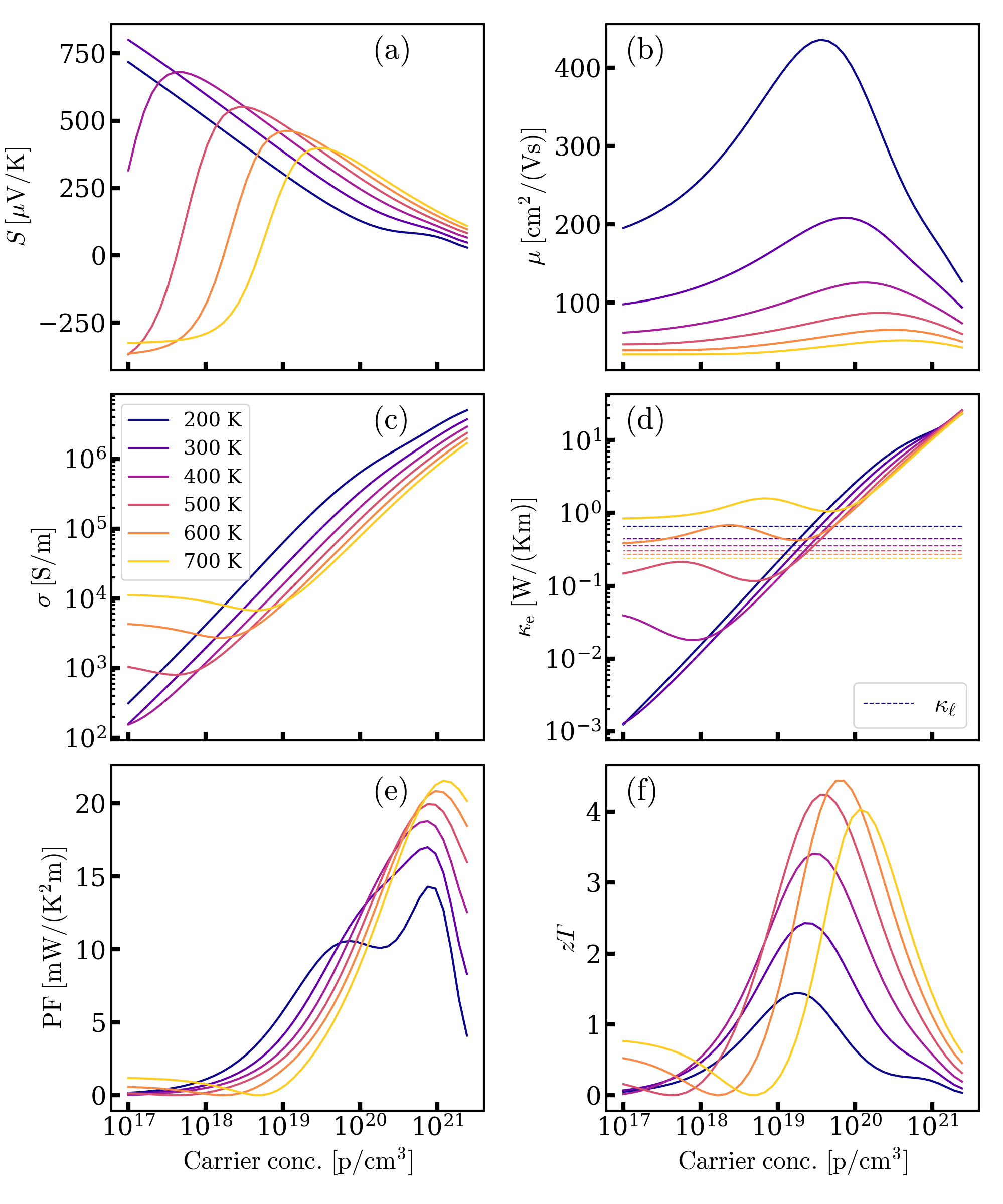}
    \caption{The Seebeck coefficient ($S$), mobility ($\mu$), electrical conductivity ($\sigma$), electron thermal conductivity ($\kappa_\mathrm{e}$), power factor (PF), and figure of merit ($zT$) of p-type Na$_2$TlSb at temperatures 200--700\;K.
    \label{fig: transport_p}}
\end{figure}

% p-type
Figure \ref{fig: transport_p} shows TE transport properties at different temperatures as a function of p-carrier concentration. With $zT$ surpassing a value of 4.4 at 600\;K, we confirm the earlier identification of  Na$_2$TlSb as an extremely promising thermoelectric material \cite{Yue_2023}. The combination of high $S$ and $\sigma$ yields a maximum PF of more than 15\;mW/K$^2$m for most temperatures. Combined with the ultra-low $\kappa_\ell$, this gives a promising material at high temperatures, but it also reaches an exceptional optimal p-type $zT$ of 2.4 at room temperature (300\;K). The fact that $zT$ remains high across such a wide temperature span can be understood from the high low-temperature mobility $\mu$. The mobility decreases markedly with temperature, due to the decreasing electron-phonon scattering at lower temperature, as discussed in Sec.~\ref{sec: scattering}. A key factor limiting $zT$ is the electronic thermal conductivity $\kappa_\mathrm{e}$ reaching as high as 0.85\;W/mK at optimal doping concentration, which is larger than $\kappa_\ell$. 

An interesting aspect of the large asymmetry between the valence and conduction DOS is that charge neutrality biases $\varepsilon_{\mathrm{F}}$ towards the conduction band, and quite large p-doping is needed to overcome bipolar transport. This is seen at 700$\;$K, where $S$ changes sign around 5$\times$10$^{19}$\;h/cm$^{3}$ (a). Similarly, both $\sigma$ and $\kappa_\mathrm{e}$ flatten out at low carrier concentrations as the conduction band begins to contribute to transport. 

\begin{figure}[h!]
    \centering
    \includegraphics[width=\linewidth]{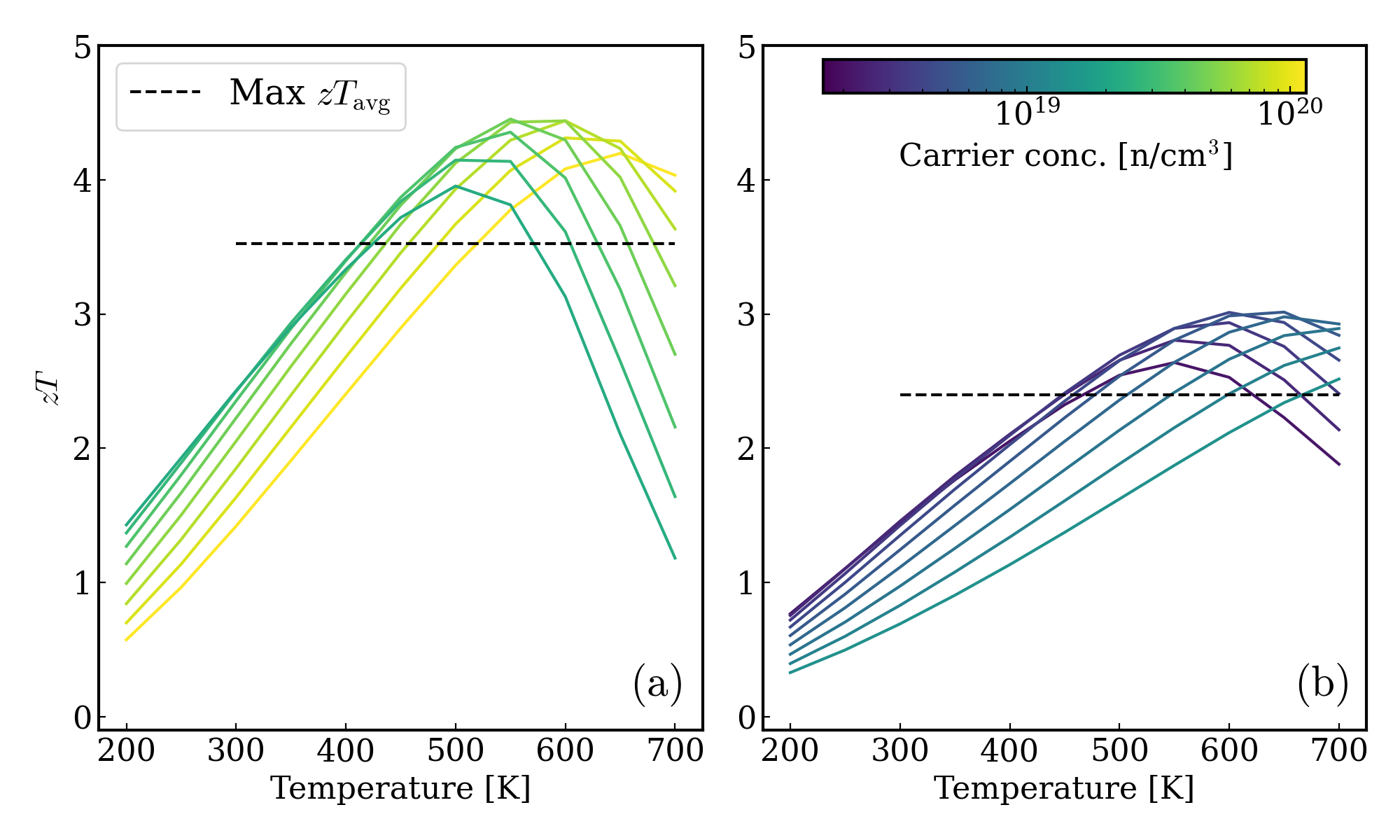}
    \caption{$zT$ as a function of temperature for various carrier concentrations for (a) p- and (b) n-type doping. The highest achievable average $zT$ over the range 300--700\;K is also indicated.
    \label{fig: zT_T}}
\end{figure}

% zT vs T
Fig.~\ref{fig: zT_T} displays (a) the p-type and (b) n-type $zT$ as a function of temperature over the range 200--700\;K. Each is plotted for different carrier concentrations around the optimal one for $zT$ ($N_\mathrm{opt}$). A consequence of the high $zT$ at a wide range of temperatures is that the temperature-averaged $zT$ can also be very high. This is shown by the dashed lines, which indicate the maximal temperature-averaged $zT$  ($zT_\mathrm{avg}$) from 300 to 700\;K, using a fixed doping concentration of 5.7$\times 10^{19}$\;h/cm$^{3}$ and  4.3$\times 10^{18}$\;e/cm$^{3}$ for p- and n-type doping, respectively. Even higher values would be obtained with graded doping in the TE legs. The high n-type $zT$, which arises from fast electrons around the L-valley, also highlights the promise of Na$_2$TlSb for single-material thermoelectric devices.

\subsection{Electronic scattering rates}
\label{sec: scattering}

\begin{figure}[h!]
    \centering
    \includegraphics[width=\linewidth]{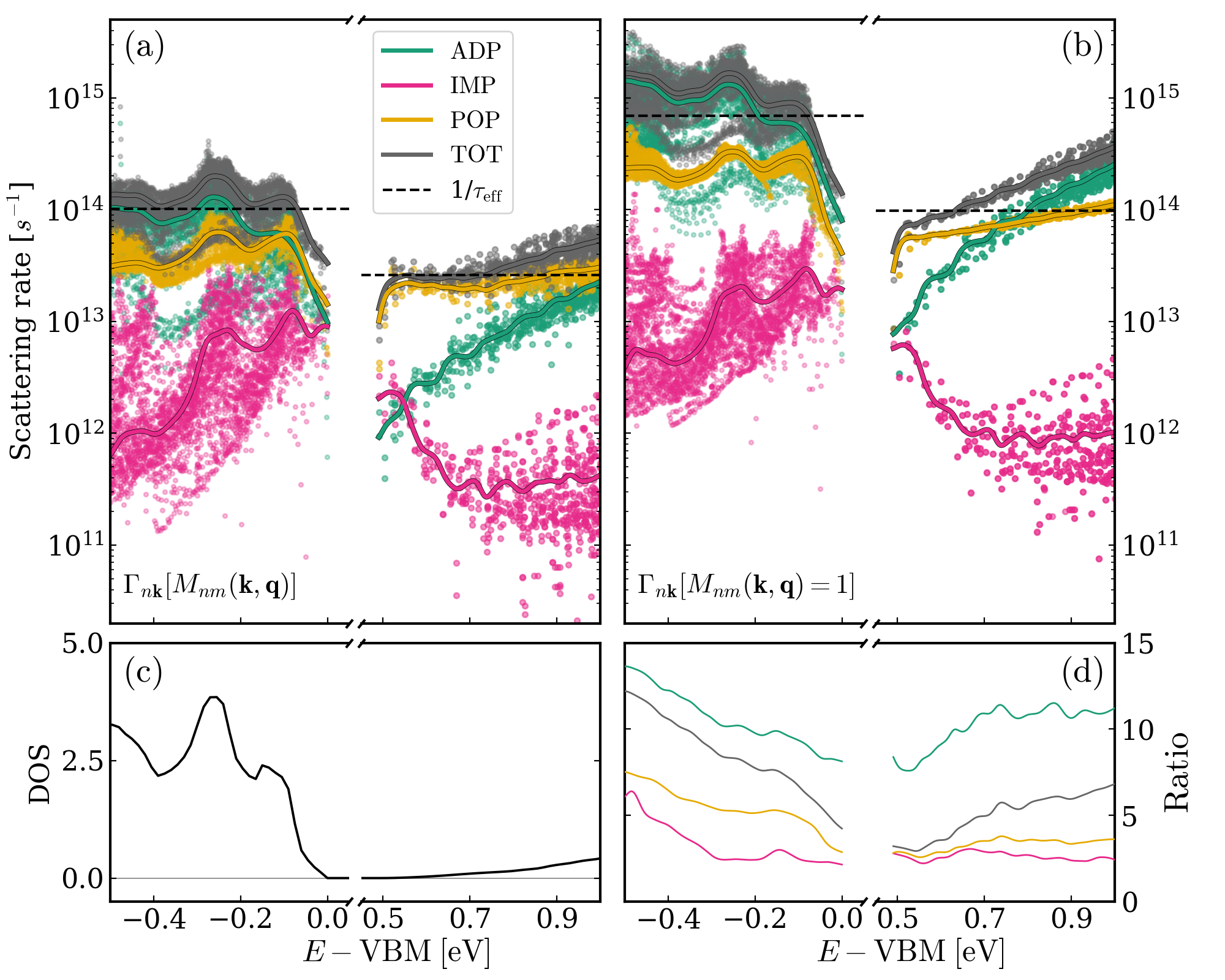}
    \caption{Scattering rates of the valence and conduction bands of Na$_2$TlSb based on ADP, IMP, and POP scattering, as well as the total scattering rate (TOT) at 600\;K. The solid lines show a Gaussian-smeared average of each scattering mechanism. In (a), the effect of wavefunction overlaps is included, while in (b), the overlap is set to unity for all transition rates. The electronic DOS is shown in (c). In (d), the ratio between the running average scattering with (from a) and without (from b) the effect of wavefunction overlap is displayed.
    \label{fig: rates}}
\end{figure}

% Scattering rates
Figure \ref{fig: rates} (a) displays the scattering rates $\Gamma_{n\mathbf{k}}$ at 600\;K with respect to the charge carrier energy for the corresponding optimal n- and p-type doping, $N_\mathrm{opt}$. Their wide scatter reflects a significant $\mathbf{k}$-dependence in scattering rates. For clarity, running averages (Gaussian smeared) are also shown. As expected from the significantly larger DOS, p-type scattering is larger than n-type. This can be seen by the dashed lines indicating 1/$\tau_\mathrm{eff}$, i.e., the effective constant scattering rate that reproduce the mobility of the full \textsc{AMSET} calculations\cite{Grimenes_2025}. Despite the huge DOS, shown in (c), the effective relaxation time of the p-doped material at 600\;K is quite similar to the standard CRTA choice of 10\;fs. For both p- and n-type, POP is the strongest scattering mechanism close to the band edge. For p-type, ADP becomes dominant further into the band. This is not the case for n-type, however, where ADP scattering is much smaller. Similarly, IMP scattering is larger for p-type compared to n-type, but it remains a minor contribution to the total scattering rate.

\begin{figure*}[t]
    \centering
    \includegraphics[width=0.7\linewidth]{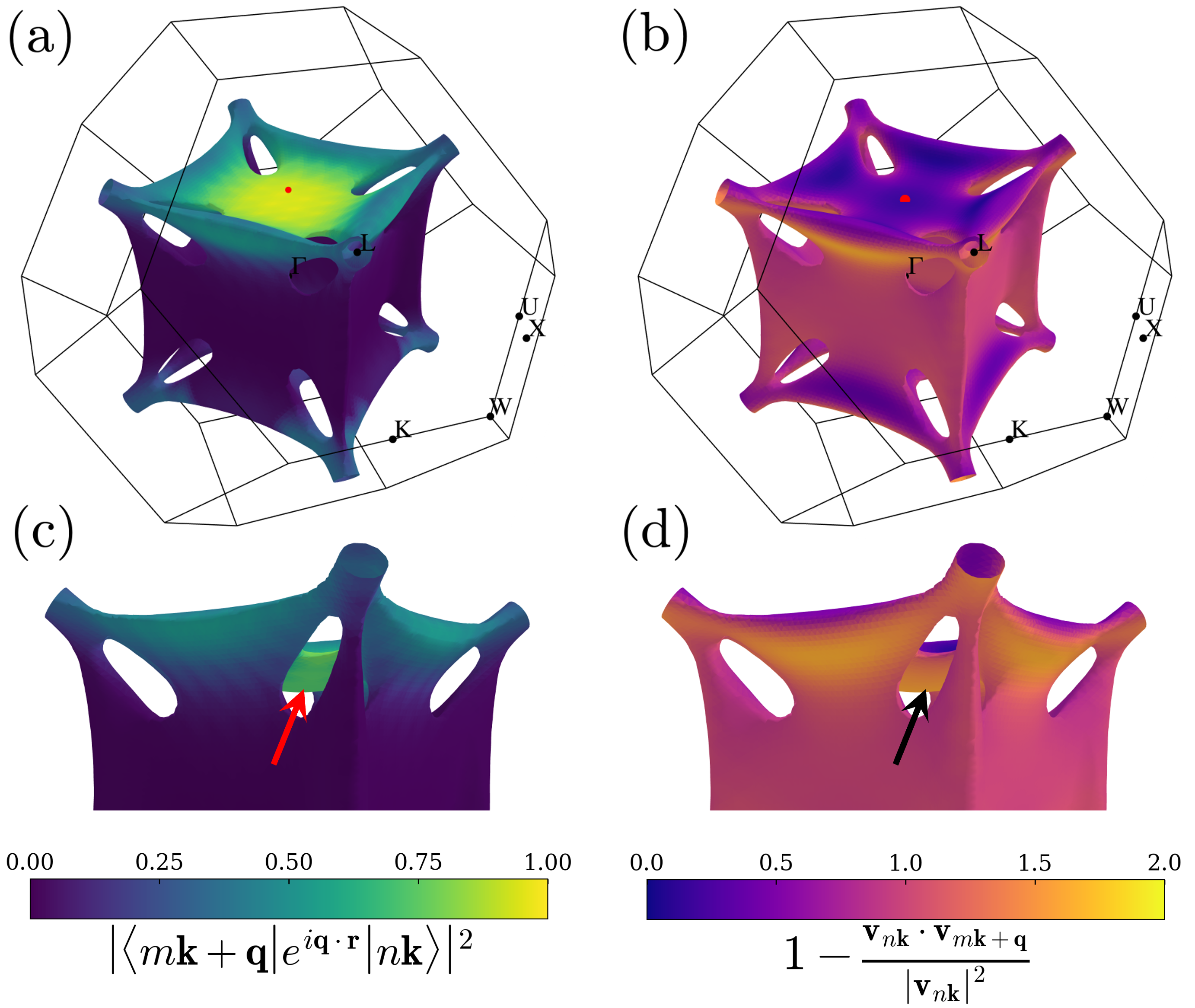}
    \caption{Energy isosurfaces of the Na$_2$TlSb valence band at 0.12\;eV below the VBM. (a) Colored by the wavefunction overlap between initial and final state $|M_{mn}(\mathbf{k},\mathbf{q})|^2 =|\langle m \mathbf{k}+\mathbf{q}|e^{i\mathbf{q}\cdot \mathbf{r}}|n\mathbf{k}\rangle|^2$ (Eq.~\ref{eq: M}), from a selected initial $\mathbf{k}$-point marked with a red dot on the isosurface. (b) Similarly colored by the geometric MRTA factor $\Lambda_{nm}(\mathbf{k},\mathbf{q})$, see Eq. \ref{eq: MRTA_factor}. Panels (c) and (d) show the same surfaces from a different perspective, revealing the inner surface. The figures are generated with IFermi \cite{Ganose_2021a}.
    \label{fig: overlap}}
\end{figure*}

% ADP and overlap
ADP scattering rates for different energies reflect the curve shape of the DOS, shown in \ref{fig: rates} (c), and are far larger for p-type than n-type (deformation potentials are comparable, see Ref.\ \cite{SM}). This trend could be expected from the lack of an explicit $\mathbf{q}$-dependence for the perturbing potential of ADP scattering (see \ref{eq: ADP1}), allowing for potentially large scattering rates across the Brillouin zone. 

However, $\mathbf{q}$-dependence enters through the wavefunction overlap $M_{mn}(\mathbf{k},\mathbf{q})= \langle m \mathbf{k}+\mathbf{q}|e^{i\mathbf{q} \cdot \mathbf{r}}|n\mathbf{k}\rangle$, as well as through the MRTA factor $\Lambda$, as shown in Fig. \ref{fig: overlap}. Panels (a) and (c) show that the wavefunction overlaps remain quite large on the upper side of the box-like isosurface, but the overlap with the other sides is nearly zero. The low overlap can be understood from the dominant p-character of the valence band (Fig.~\ref{fig: band_structure}), i.e., wavefunctions on different sides are (approximately) orthogonal. The effect of this can be illustrated by artificially fixing the overlap to $M=1$: it increases the scattering rates by a factor of 7 for p-type and 4 for n-type, as shown in Fig.~\ref{fig: rates}. Comparing panel (a) and (b) (and the ratio in (d)) shows that setting $M=1$ has a far larger effect on ADP-scattering than for the other scattering mechanisms.

% MRTA
The geometric MRTA factor $\Lambda_{nm}(\mathbf{k},\mathbf{q}) = 1-\mathbf{v}_{n\mathbf{k}} \cdot \mathbf{v}_{m\mathbf{k}+\mathbf{q}} /|\mathbf{v}_{n\mathbf{k}}|^2$ (Eq. \ref{eq: MRTA_factor}) for elastic scattering is shown projected onto the energy isosurface in Fig.~\ref{fig: overlap} (b/d). This factor has values around $0$ for forward scattering, $1$ for perpendicular scattering, and $2$ for backwards scattering. The box-like geometry of the Na$_2$TlSb energy isosurface makes this geometric factor very evident. It is very low for same-side scattering, where $M$ is also largest, and has higher values on the other faces, where $M$ is low. However, on the inside of the box (emphasized by the arrows), both  $M$ and $\Lambda$ are large. We find that the effect of including the MRTA factor is small for the ADP scattering but large for IMP, as detailed in Fig.~S4 in Ref.~\cite{SM}, reflecting the short and long nature of these scattering mechanisms.
 
\begin{figure}[h!]
    \centering
    \includegraphics[width=\linewidth]{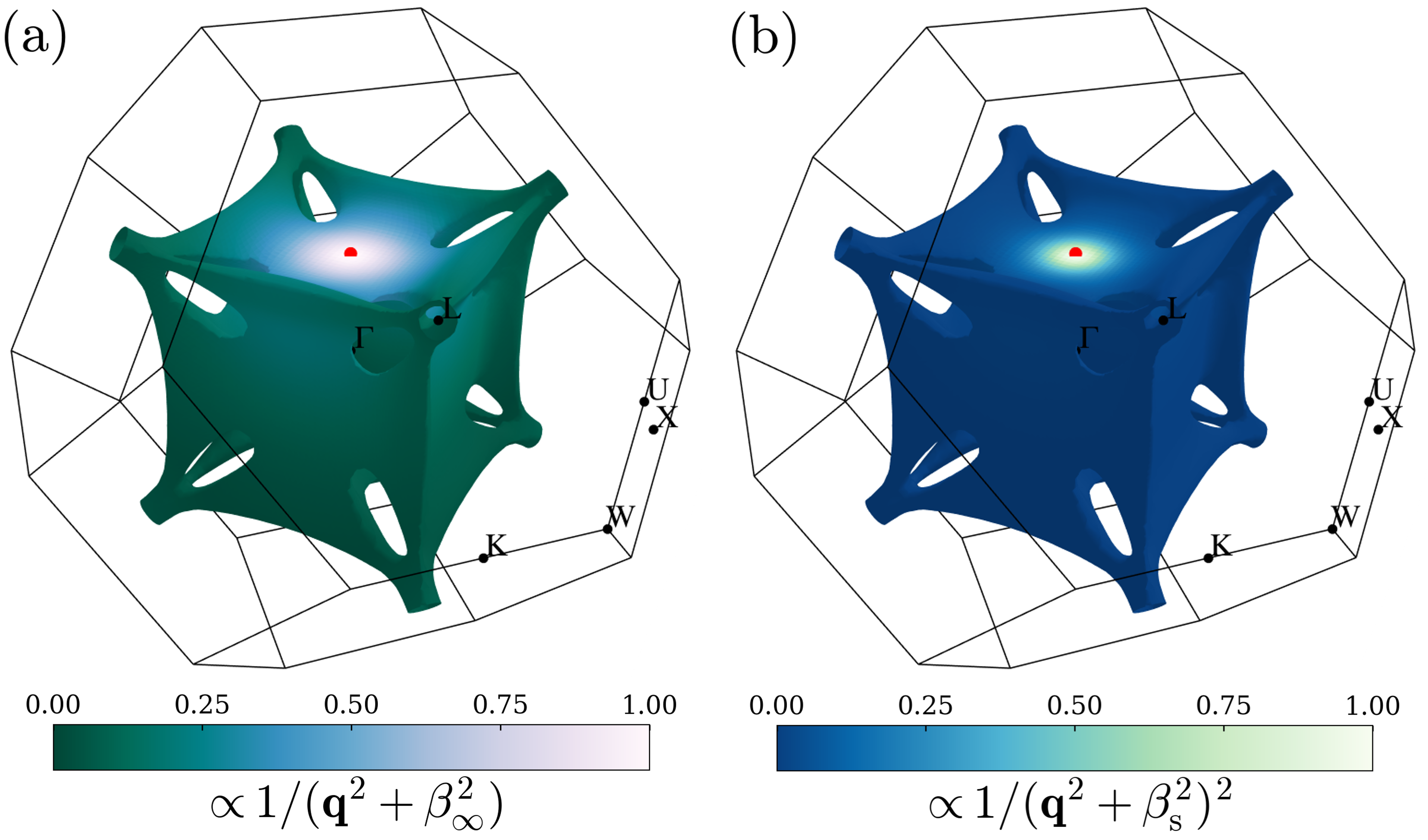}
    \caption{Energy isosurfaces of the Na$_2$TlSb valence band at 0.12\;eV below VBM. (a) Colored by the normalized q-dependence 1/$(\mathbf{q}^2 + \beta^2_\infty)$ for POP scattering and (b) the normalized q-dependence 1/$(\mathbf{q}^2 + \beta^2_\mathrm{s})^2$ for IMP scattering from a sample $\mathbf{k}$-point (red dot).
    \label{fig: q_dist}}
\end{figure}

% q-distance and free-carrier screening for POP and IMP
Fig.~\ref{fig: q_dist} shows the contributions of the explicit $\mathbf{q}$-dependent factors in POP (a, Eq.~\ref{eq: POP1}) and IMP (b, Eq.~\ref{eq: IMP1}) scattering. It underlines how these scattering mechanisms are strongly localized in $\mathbf{k}$-space, and can in part explain why POP scattering rates have similar n-type and p-type magnitudes despite much large energy isosurfaces on the p-side. The increased scattering due to larger energy isosurfaces is also counteracted by increased screening (Eq. \ref{eq: POP1}) at optimal carrier concentrations $N_\mathrm{opt}$, which is much higher when the DOS is large. The $\beta_\infty^2$ term of p-type is 17 times higher than that of n-type for their respective $N_\mathrm{opt}$. The effect of POP scattering is provided in the supplementary material (SM) \cite{SM}. Interestingly, without free-carrier screening, the p-type $zT_\mathrm{avg}$ is only marginally reduced since the increased scattering reduces bipolar transport at higher temperatures (details in \cite{SM} Fig. S1). While p-type IMP scattering is even more local, it remains larger than for the n-type Fermi pockets, in part this can be attributed to the high $N_\mathrm{opt}$ for p-type, making the prefactor (eq. \ref{eq: IMP}) larger. Even if POP and IMP scattering are short-ranged, they can be very strong, particularly close to the band edges (Fig.~\ref{fig: rates}) due to large scattering prefactors and divergent $1/q$ behavior in the small $\mathbf{q}$ limit.

\subsection{Comparison with ab-initio electron-phonon scattering}
\begin{figure}[h!]
    \centering
    \includegraphics[width=\linewidth]{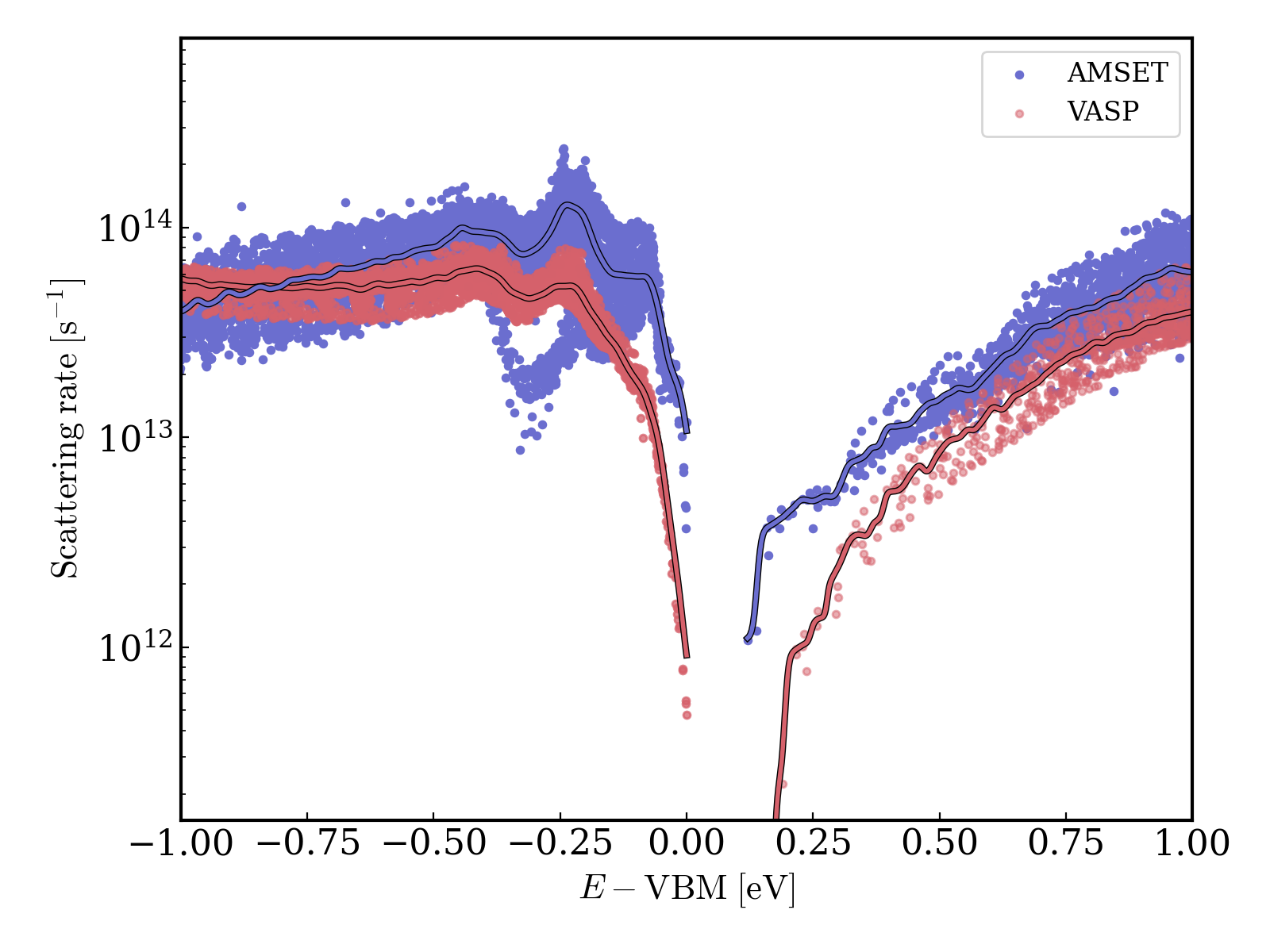}
    \caption{Electron scattering rates based on electron-phonon interactions calculated from first principles in \textsc{VASP} compared to phonon-based (ADP and POP) scattering rates obtained with \textsc{AMSET}. The scattering rates are calculated at 600\;K and with $N_\mathrm{c}=7.2\times10^{19}$\;h/cm$^{3}$ and $4.3\times10^{18}$\;e/cm$^{3}$ for the valence and conduction band, respectively. Running averages are shown as solid curves.
    \label{fig: vasp_rates}}
\end{figure}

Figure \ref{fig: vasp_rates} presents the results of the \textsc{AMSET} benchmarking, as detailed in Sec.~\ref{sec: benchmark}. It compares ADP+POP scattering rates from \textsc{AMSET} with the full ab initio electron-phonon scattering rates computed with \textsc{VASP}. Compared to \textsc{VASP}, \textsc{AMSET} systematically overestimates scattering rates by 1.5--3 times for most of the displayed energies, suggesting that \textsc{AMSET} overestimates phonon-based scattering. The difference is largest close to the band edge. Still, there is good agreement between \textsc{VASP} and \textsc{AMSET} in terms of the energy dependence of the scattering rates.

\section{Discussion}
A concern for materials with high DOS is that excessive electron scattering can cripple the electron mobility, in line with the numerous tradeoffs limiting the development of high-performing TE materials \cite{Park_2021a}. However, in our \textsc{AMSET}-based analysis, we found this not to be the case for Na$_2$TlSb, an extreme example of extended 2D-like energy isosurfaces akin to those of 1D-like nanostructured materials. For IMP and POP scattering, the momentum-dependence of the perturbing potential combined with large free-carrier screening at $N_\mathrm{opt}$ reduced the scattering rates. ADP scattering lacks these mechanisms and is the largest, but its magnitude is also curbed by the wavevector dependence of the wavefunction overlaps.

The disparity between the full \textsc{VASP}-based electron-phonon calculations in the benchmark study underlines the approximative nature of \textsc{AMSET}. Nonetheless, the fact that scattering rates with \textsc{VASP} were consistently lower for all relevant energies indicates that that transport calculations at the same theoretical level as for our main results (hybrid functionals, completely converged numerical parameters, etc.), would result in a $zT$ even higher the value presented above.

It is critical to note that other mechanisms not accounted for here could also change the scattering rates. Electron-electron scattering, which is usually insignificant for lightly doped semiconductors, could become significant at carrier concentrations around the optimal p-type, $N_\mathrm{opt}$ \cite{Fischetti_1991}. The large scattering space might also give rise to significant multi-phonon electronic scattering rates \cite{Lee_2020}. The use of simple Thomas-Fermi-based free-carrier screening might also be inadequate at such large carrier concentrations \cite{Go_2025}. It is also worth noting that an experimental material system would exhibit additional scattering mechanisms, such as grain-boundary and point-defect scattering. Finally, the relaxation time approximation has for some systems been found to deviate significantly from the more sophisticated iterative approaches to solving the Boltzmann transport equation \cite{Claes_2022,Claes_2025}.

Concerns can also be raised regarding the realizability of Na$_2$TlSb. Yeu \textit{et al}.\ \cite{Yue_2023} found it to be stable at 700\;K with \textit{ab initio} molecular dynamics, but the material has yet to be synthesized to the best of the authors' knowledge. Materials containing alkali metals, such as Na, generally require more care during synthesis due to high reactivity with oxygen and water, and tend to suffer from degradation in use as a TE material at high temperature due to ionic diffusion. In addition, the high doping needed to reach the p-type $N_\mathrm{opt}$ could be difficult to achieve without significantly altering other properties of the material. On the other hand, it has been argued that materials with a high $N_\mathrm{opt}$ are less vulnerable to small changes in $N$ stemming from impurities in the precursors, thereby reducing cost \cite{Lou_2024}. Finally, the toxicity of Tl will necessitate a high level of care during synthesis, in use, and when disposing of the material at the end of its lifetime.

We have used the lattice thermal conductivity numbers from Yue \cite{Yue_2023} and use the same method for calculating electronic transport, but still achieved different transport properties. The difference can be explained by some methodological choices. In our work, the HSE06 band structure was used for transport calculations, not PBEsol scissored to the band gap of HSE06  employed in the previous work, resulting in differences in the band structure. Second, the materials properties used in \textsc{AMSET} also differ; we have, e.g.\ found a significantly lower ionic contribution to the dielectric constant. Combined with the inclusion of free-carrier screening, the POP scattering rates are greatly reduced in our work. Finally, we used \textsc{AMSET} version 0.5.0, which features an improved treatment of wavefunction overlaps for scattering across Brillouin zone boundaries, a change that is likely to increase scattering rates.

Despite the practical issues of realizing this material, our study represents an important addition to the investigation of electron scattering in materials with low-dimensional band structures \cite{Parker_2013,Bilc_2015,Dylla_2019,Park_2021,Brod_2021,Grimenes_2025}. While we do believe that Na$_2$TlSb itself merits further investigation due to its very attractive band structure, the study foremost calls for intensifying the hunt for material structures exhibiting low-dimensional features in the near gap regions, preferably also less-toxic compounds. Many of the specific findings should carry over to low-dimensional materials of the PbTe-prototype and related Heusler alloys, due to their similar bonding nature \cite{He_2019,Brod_2020,Brod_2021}, and help formulate design principles to improve these and other materials. Critically, our study emphasizes how energy isosurfaces that are delocalized in $\mathbf{k}$-space tend to exhibit low electron scattering, makes such materials promising.

\section{Conclusion}
We have calculated the electronic transport properties of Na$_2$TlSb with scattering rates from \textsc{AMSET}. The valence band displays 1D-like features that result in a sharp increase in the DOS close to the band edge, while retaining a high mobility. Despite the high DOS, the electron scattering remains modest due to low wavefunction overlaps, large scattering distances in $k$-space, and high free-carrier screening. Combined with an ultra-low $\kappa_\ell$, this results in an exceptionally high predicted $zT$ over a wide range of temperatures. The results underline the huge potential of low-dimensional features in electronic band structures for high TE performance.

\section{Data availability}
Input files necessary to reproduce the main results in this article are available at \url{10.17172/NOMAD/2026.03.09-1}.

\section{Acknowledgment}
Ø.A.G., O.M.L., and K.B.\ are supported by the Research Council of Norway through the Allotherm project (Project No. 314778). Computational resources were provided by the Norwegian e-infrastructure for research and education, Sigma2, through grant No.\ nn9711k. We acknowledge insightful discussions with L. Chaput and G.J. Snyder.

\appendix

\section{Lattice thermal conductivity}
\label{app: ltc}
Table \ref{tab: ltc} lists the $\kappa_\ell$ used for calculations of $zT$. The numbers on the left are extracted from Ref.\ \cite{Yue_2023}, while the temperatures in between (right side) are interpolated with cubic splines. 

\begin{table}[h!]
  \caption{Lattice thermal conductivity values.}
\label{tab: ltc}
\begin{ruledtabular}
\begin{tabular}{lllllll}
$T$ [K] & $\kappa_\ell$ [Wm$^{-1}$K$^{-1}$] & $T$ [K] & $\kappa_\ell$ [Wm$^{-1}$K$^{-1}$] \\
\hline & \\[-1.5ex]
200    & 0.65      & 250   & 0.53  \\
300    & 0.44      & 350   & 0.38  \\
400    & 0.35      & 450   & 0.32  \\
500    & 0.30      & 550   & 0.28  \\
600    & 0.27      & 650   & 0.26  \\
700    & 0.24      \\

\end{tabular}
\end{ruledtabular}
\end{table}

\section{n-type electron transport properties}
\label{app: n_type}

\begin{figure}[h!]
    \centering
    \includegraphics[width=\linewidth]{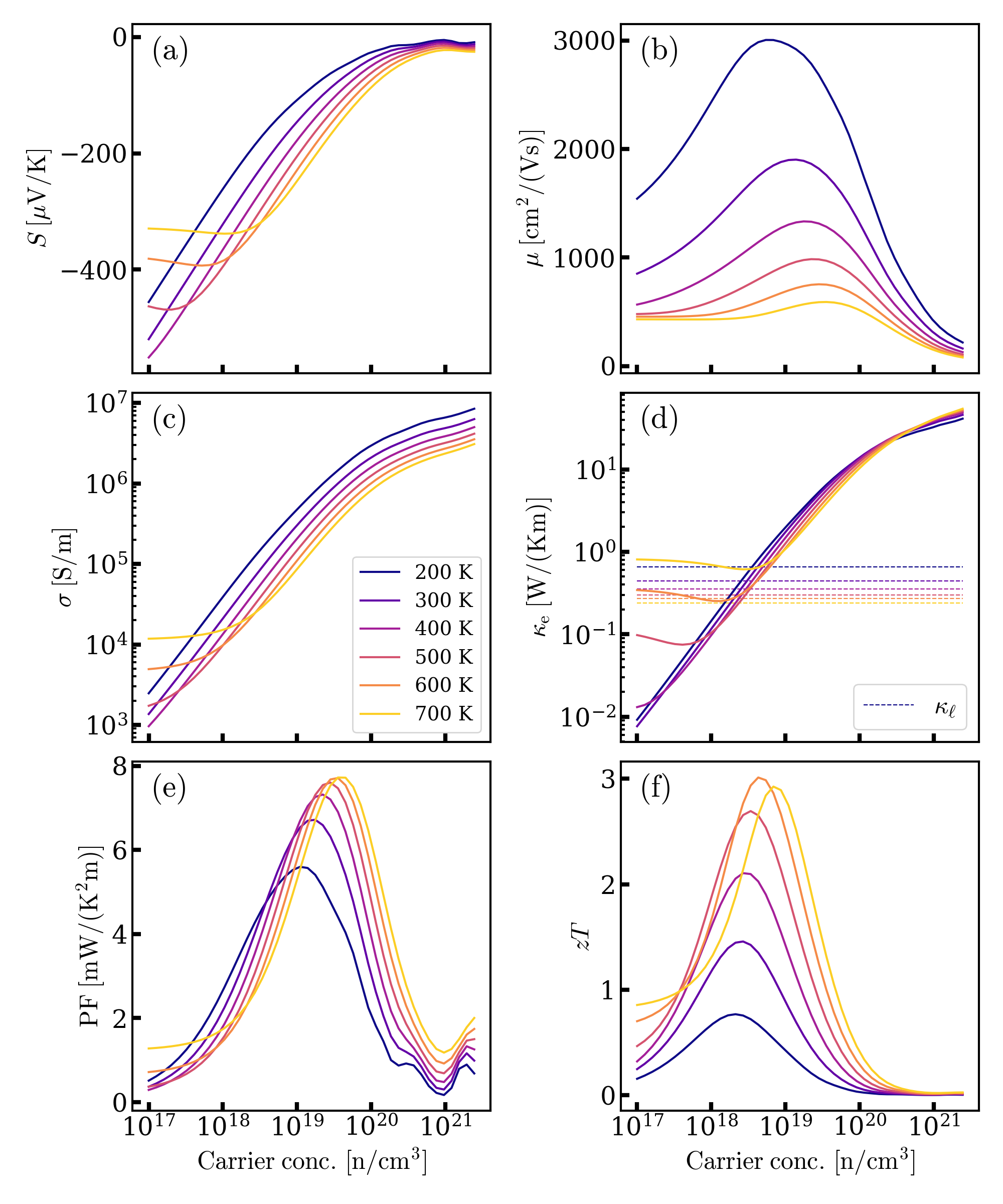}
    \caption{The Seebeck coefficient, electrical conductivity, electron thermal conductivity, power factor, and figure of merit of n-type Na$_2$TlSb at temperatures 200--700\;K.
    \label{fig: transport_n}}
\end{figure}

% n-type
The n-type electron transport properties of Na$_2$TlSb at temperatures 200--700$\;$K are shown in Fig.~\ref{fig: transport_n}. The highly dispersive conduction band, with a high relaxation time, gives a $\mu$ around an order of magnitude larger than that of the p-type. However, the lower DOS results in $S$ more quickly approaching zero in the high-doping regime, thereby limiting the PF. We still predict an impressive $zT$ ranging from 1.5 at 300$\;$K to 3.0 at 700$\;$K. Interestingly, at 700\;K the difference in effective mass and mobility results in a large negative $S$ at low carrier concentration. As a result, Na$_2$TlSb has an n-type $zT$ of 0.85 at 700\;K without doping.

\bibliography{ref, kb}

\end{document}

\title{Exceptional thermoelectric properties in Na$_2$TlSb enabled by quasi-1D band structure}

\author{Øven A. Grimenes}
\affiliation{
 Department of Mechanical Engineering and Technology Management, \\ Norwegian University of Life Sciences, NO-1432 Ås, Norway}

\author{Ole M. Løvvik}
\affiliation{SINTEF Sustainable Energy Technology, Forskningsveien 1, NO-0314 Oslo, Norway}

\author{Kristian Berland}
\affiliation{
 Department of Mechanical Engineering and Technology Management, \\ Norwegian University of Life Sciences, NO-1432 Ås, Norway}

\date{\today}

\maketitle

\section{Free-carrier screening}
\begin{figure}[h!]
    \centering
    \includegraphics[width=0.7\linewidth]{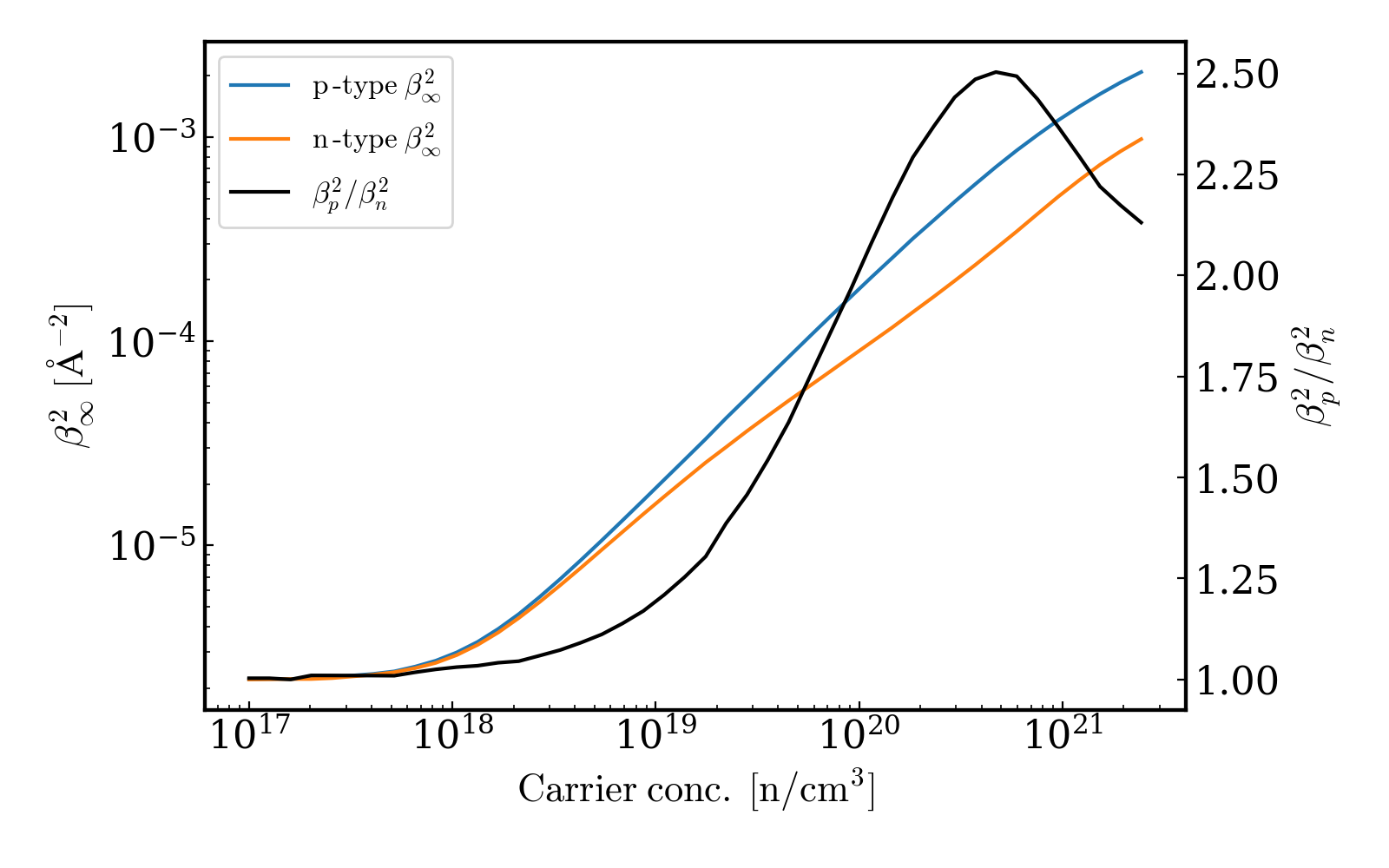}
    \caption{Inverse high-frequency free-carrier screening length for p- and n-type Na$_2$TlSb at 600\;K for various carrier concentrations and the ratios between them.
    \label{fig: screening}}
\end{figure}

\newpage

\section{Deformation potentials}
\begin{figure}[h!]
    \centering
    \includegraphics[width=0.9\linewidth]{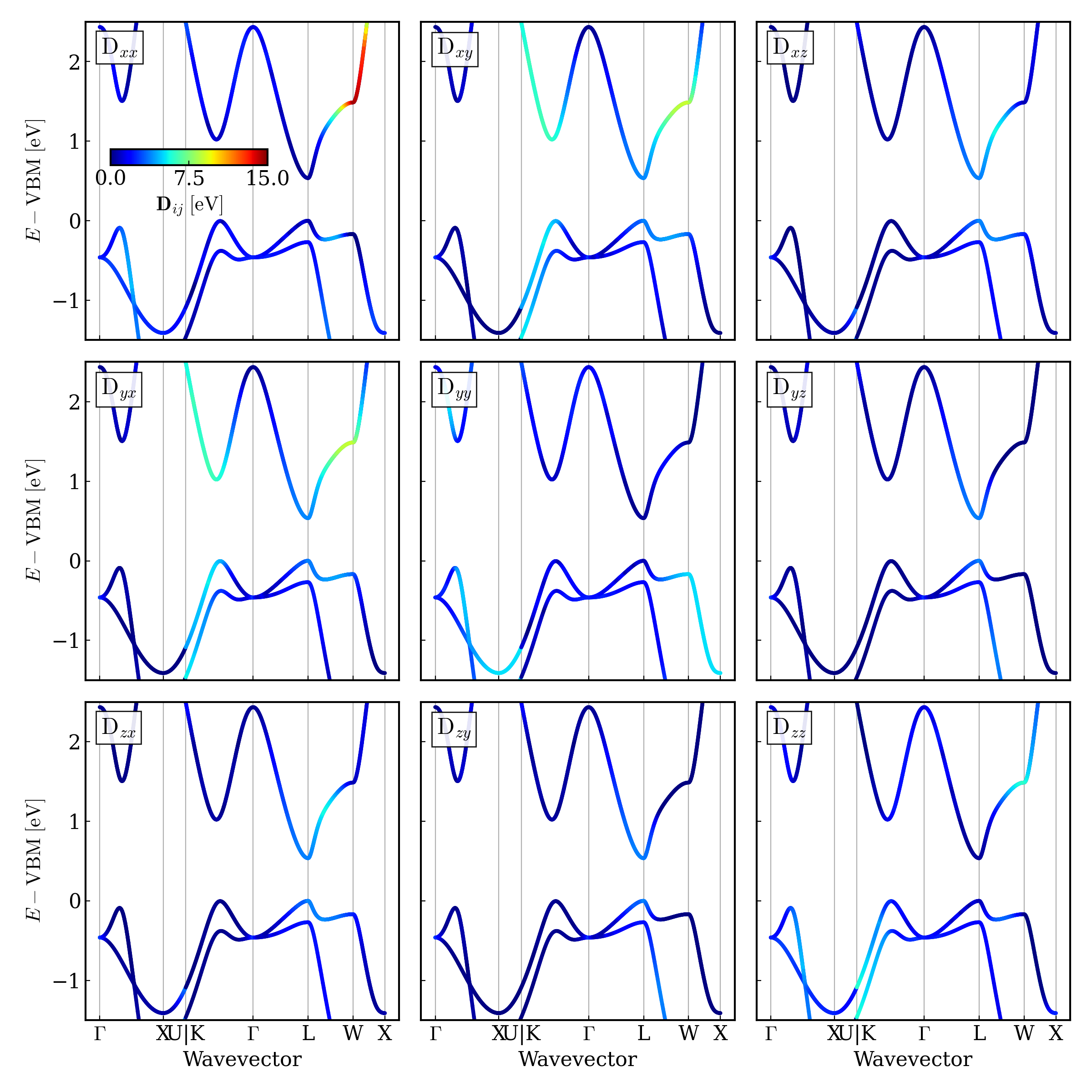}
    \caption{Tensor components of the deformation potential of Na$_2$TlSb projected onto the band structure.
    \label{fig: deform_pot}}
\end{figure}

\newpage

\section{$zT$ without free-carrier screening}
\begin{figure}[h!]
    \centering
    \includegraphics[width=0.9\linewidth]{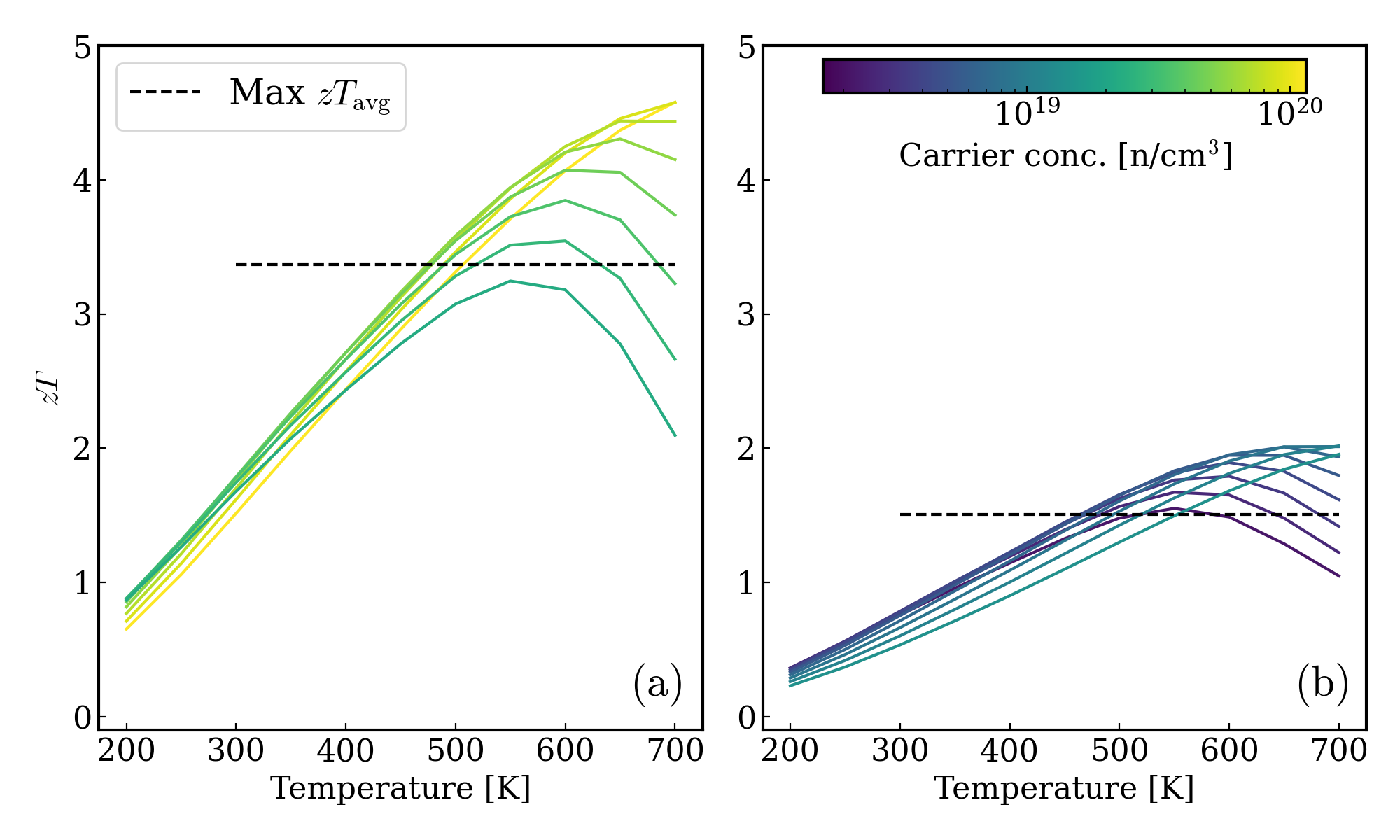}
    \caption{p- and n-type $zT$ of N2$_2$TlSb as a function of temperature without inclusion of free-carrier screening. The $zT_\mathrm{avg}$ of p-type is almost as high as with free-carrier screening included due to reduced bipolar transport at high temperatures.
    \label{fig: no_screening}}
\end{figure}

\newpage

\section{ADP and IMP without MRTA-factor}
\begin{figure}[h!]
    \centering
    \includegraphics[width=0.9\linewidth]{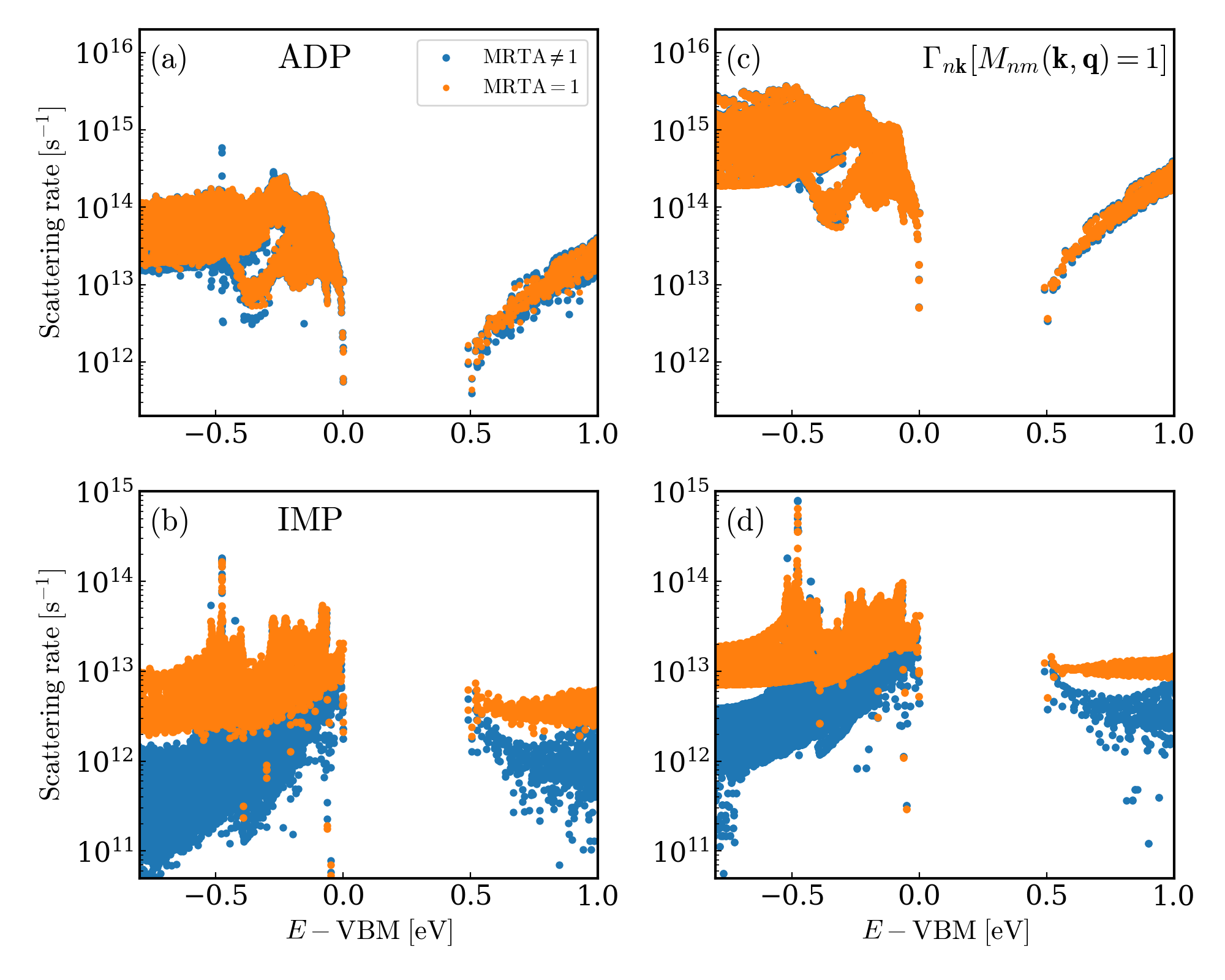}
    \caption{ADP (a)/(c) and IMP (b)/(d) scattering rates at 600\;K with and without the MRTA factor. In the column on the right the wavefunction overlap is set to unity. 
    \label{fig: phonon}}
\end{figure}

\newpage

\section{Phonon dispersion}
\begin{figure}[h!]
    \centering
    \includegraphics[width=0.9\linewidth]{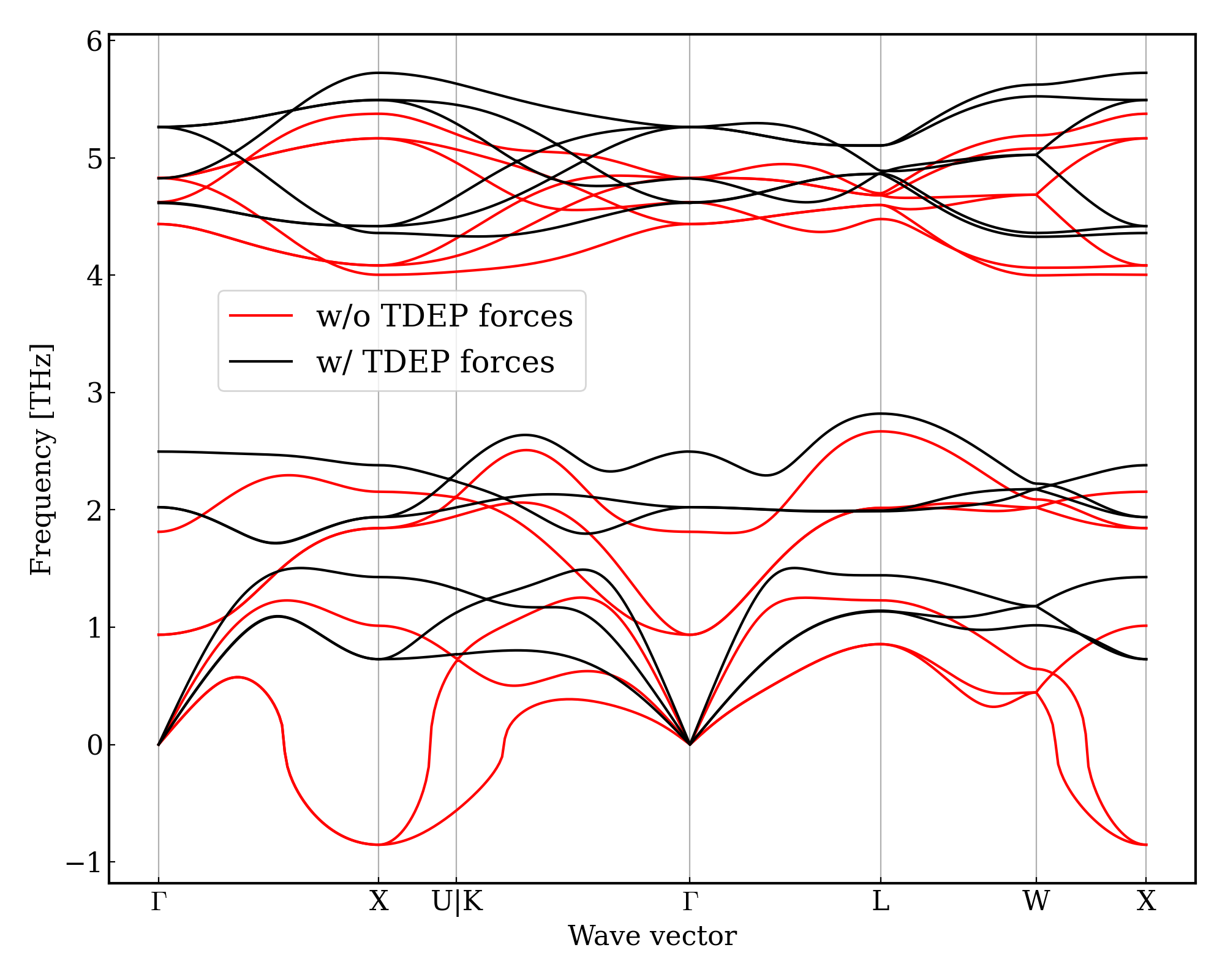}
    \caption{Phonon dispersion of Na$_2$TlSb calculated with finite differences using \textsc{Phelel} (red), and with \textsc{sTDEP} (black).  
    \label{fig: phonon}}
\end{figure}